\documentclass[aps,prc,twoside,twocolumn,nofootinbib,10pt,showpacs,floatfix]{revtex4-1}
\usepackage{amsmath,amssymb}
\usepackage{graphicx,bm}
\usepackage{slashed}
\usepackage{epstopdf}
\usepackage{ulem} %% for strike-through
\usepackage[usenames]{color}
\usepackage{float}
\usepackage{hyperref}
\usepackage{subfigure}
\usepackage{rotating}
\usepackage{color}
\usepackage{multirow}
\usepackage{dcolumn}
\usepackage{overpic}
\usepackage{booktabs}
\usepackage{makecell}
\usepackage{diagbox}
\usepackage{array}
\newcolumntype{|}{!{\vline}}

\renewcommand\sout{\bgroup \color{red} \ULdepth=-.5ex \ULset}

\newsavebox{\tablebox}
\begin{document}
%%%%%%%%%%%%%%%%%%%%%%%%%%%%%%%%%%%%%%%%%%%%%%%%%%%%%%%%%%%%%%%%%%%
\title{Doubly charmed dibaryon states}
%%%%%%%%%%%%%%%%%%%%%%%%%%%%%%%%%%%%%%%%%%%%%%%%%%%%%%%%%%%%%%%%%%%%

\author{Zhe Liu$^{1,2}$}\email{zhliu20@lzu.edu.cn}
\author{Hong-Tao An$^{1,2}$}\email{anht14@lzu.edu.cn}
\author{Zhan-Wei Liu$^{1,2,3}$}\email{liuzhanwei@lzu.edu.cn}
\author{Xiang Liu$^{1,2,3}$}\email{xiangliu@lzu.edu.cn}
\affiliation{
$^1$School of Physical Science and Technology, Lanzhou University, Lanzhou 730000, China\\
$^2$Research Center for Hadron and CSR Physics, Lanzhou University and Institute of Modern Physics of CAS, Lanzhou 730000, China\\
$^3$Lanzhou Center for Theoretical Physics, Key Laboratory of Theoretical Physics of Gansu Province, and Frontiers Science Center for Rare Isotopes, Lanzhou University, Lanzhou 730000, China}

\date{\today}
%%%%%%%%%%%%%%%%%%%%
\begin{abstract}
%%%%%%%%%%%%%%%%%%%%%%%%%%%%%%%%%%%%%%%%%%%%%%%%%%%%%%%%%%%%%%%%%%%%

In this work, we study the doubly charmed dibaryon states with the $qqqqcc$ ($q =u, d, s$) configuration. 
The mass spectra of doubly charmed dibaryon states are obtained systematically within the chromomagnetic interaction model. 
In addition to the mass spectrum analysis, we illustrate their two-body strong decay behaviours. 
Our results suggest that there may be narrow states or even stable states that cannot decay through the strong interaction. 
We hope that our results will provide valuable information for further experimental searches for doubly charmed dibaryon states.

%%%%%%%%%%%%%%%%%%%%%%%%%%%%%%%%%%%%%%%%%%%%%%%%%%%%%%%%%%%%%%%%%%%%
\end{abstract}
%%%%%%%%%%%%%%%%%%%%%%%%%%%%%%%%%%%%%%%%%%%%%%%%%%%%%%%%%%%%%%%%%%%%

\maketitle
\thispagestyle{empty} %

%%%%%%%%%%%%%%%%%%%%%%%%%%%%%%%%%%%%%%%%%%%%%%%%%%%%%%%%%%%%%%%%%%%%
\section{Introduction}
\label{Sec:Introduction}
%%%%%%%%%%%%%%%%%%%%%%%%%%%%%%%%%%%%%%%%%%%%%%%%%%%%%%%%%%%%%%%%%%%%

With the accumulation of high-precision experimental data, the search for
exotic hadronic matter is entering a new phase. In the last two decades, numerous of charmoniumlike $XYZ$ states, $P_c$ states, and the double charm $T_{cc}^+$ state have been reported experimentally, stimulating extensive studies on exotic hadron spectroscopy (see review articles \cite{Chen:2016qju,Hosaka:2016pey,Liu:2019zoy,Brambilla:2019esw,Chen:2022asf} for details).
Recently, the LHCb Collaboration announced new observations of the $T_{c\bar{s}0}^a(2900)^{++,0}$ state in the processes $B^0\to \bar{D}^0D_s^+\pi^-$ and $B^-\to \bar{D}^-D_s^+\pi^+$ \cite{LHCb:2022026,LHCb:2022027}, and a $P_{\psi s}^{\Lambda}(4338)$ state in $B^-\to J/\psi \Lambda \bar{p}$ \cite{LHCb:2022jad}.
In addition, the CMS Collaboration confirmed the observation of the $X(6900)$ state and found more structures in the di-$J/\psi$ invariant mass spectrum \cite{CMS:2022di}, and the ATALAS Collaboration confirmed the discovery of the $P_c$ states and $X(6900)$ \cite{Atalas:2022pc}.
Clearly, we are entering a new era of ``Particle Zoo 2.0," which may deepen our understanding of the non-perturbative behaviour of the strong interaction.

Among the various exotic hadrons, the hexaquark or dibaryon state is a special group.
The search for the dibaryon has a long history with many twists and turns.
One can find that a review of the long-standing search for the dibaryon system, from the early days until the last few years \cite{Clement:2016vnl}. 
The best known dibaryon state is the deuteron, which was discovered in 1932 by Urey, Brickwedde, and Murphy \cite{Harold:1932ah}.
It is a loosely bound state of a proton and neutron with a quark content of $uuuddd$, $I = 0$, $J = 1$.

Another widely believed non-strange dibaryon candidate is $d^{*} (2380)$.
Dyson and Xuong first predicted a well-known candidate $\Delta\Delta$ resonance state based on the SU(6) symmetry of the strong interaction \cite{Dyson:1964xwa}, after proposing the quark model \cite{Gell-Mann:1964ewy}.
In 1989, Goldman {\it et al.} proposed “an inevitable nonstrange dibaryon”, named $d^{*}$ because of its unique symmetry features.
About 20 years later, the $d^{*} (2380)$ resonance with the $M=2.37$ GeV, $\Gamma=70$ MeV, and $I(J^{P}) = 0(3^{+})$ was reported by the WASA-at-COSY and CELSIUS/WASA Collaborations \cite{WASA-at-COSY:2011bjg}.
It comes from the exclusive reaction channel $pn\rightarrow d\pi^{0}\pi^{0}$, and its existence was supported by later experiments \cite{WASA-at-COSY:2012uua,WASA-at-COSY:2013fzt,WASA-at-COSY:2014dmv,WASA-at-COSY:2014qkg}.
In addition, many theoretical papers properly described the properties of this resonance \cite{Buchmann:1998mi,Ping:2008tp,Huang:2013nba,Yuan:1999pg,Dai:2005kt,Bashkanov:2013cla,Gal:2013dca,Gal:2014zia,Chen:2014vha,Park:2015nha,Kim:2020rwn,Lu:2017uey,Beiming:2021bkj}.

For the strange dibaryon, Jaffe first used the MIT bag model to predict the existence of the $H$ dibaryon, which is a bound state containing the $uuddss$ quark component, in 1977 \cite{Jaffe:1976yi}.
He showed that this particle should be bound with an energy of about 70 MeV relative to the $\Lambda\Lambda$ threshold.
Meanwhile, the model study based on the Goldstone boson exchange interaction cast doubt on the existence of the $H$ dibaryon \cite{Stancu:1998ca}.
Similarly, experimental searches have ruled out the existence of such a deeply bound state  \cite{Takahashi:2001nm,Belle:2013sba}.
In addition, the $H$ dibaryon has been the subject of numerous lattice QCD calculations in recent years and the results have ruled out a deeply bound state, suggesting rather an unbound or a very weakly bound state if it exists \cite{Green:2021qol,Francis:2018qch,Luo:2011ar}.
In addition to these studies, this state had been extensively researched using various models \cite{Faessler:1982ik,Oka:1983ku,Balachandran:1983dj,Park:2016cmg,Golowich:1986chg,Yost:1985mj,Balachandran:1985fb,Mulders:1980vx}.
Another interesting candidate was introduced later by Goldman {\it et al.} \cite{Goldman:1987ma}, which is expected to appear as a $N\Omega$ resonance.
Its quark content ($uudsss$, $I = 1/2$, $J = 2$) is relatively favorable.
The STAR experiment at the Relativistic Heavy-Ion Collider (RHIC) has made significant progress in the search for the $N\Omega$ state, supporting its existence \cite{STAR:2018uho}.
Furthermore, the ALICE Collaboration at the LHC has studied the strong interactions between hadrons, providing further support for the possible formation of the $N\Omega$ state \cite{ALICE:2020mfd}.
J. Leandri and B. Silvestre-Brac conducted a series of investigations on dibaryons in the framework of a pure chromomagnetic Hamiltonian \cite{Silvestre-Brac:1992xsl, Leandri:1993zg, Leandri:1995zm}.
They showed that the inclusion of heavy quarks in the dibaryon sector can lead to configurations that are stable against decay into two baryons.
Furthermore, they proposed a number of new heavy states that could be stable under the strong interaction.
Accordingly, a more realistic way to search for the dibaryons is to turn to the heavy flavor states, where the structures are easier to detect experimentally.

First, for fully heavy dibaryons, Junnarkar {\it et al.} used a state-of-the-art lattice QCD calculation and suggested that $\Omega_{c}\Omega_{cc}$ ($sscscc$), $\Omega_{b}\Omega_{bb}$ ($ssbsbb$), and $\Omega_{ccb}\Omega_{cbb}$ ($ccbcbb$) were stable under strong and electromagnetic interactions \cite{Junnarkar:2019equ}. 
They also found that the binding of these dibaryons became stronger as their mass increased.
Lyu {\it et al.} used lattice QCD to study the $\Omega_{ccc}\Omega_{ccc}$ in the $^{1}S_{0}$ channel and found that this system is loosely bound at about 5.68 MeV \cite{Lyu:2021qsh}.
Immediately afterward, the extended one-boson exchange (OBE) model \cite{Liu:2021pdu} was used to study the existence of $\Omega \Omega$, $\Omega_{ccc}\Omega_{ccc}$, and $\Omega_{bbb}\Omega_{bbb}$ systems.
Additionally, the QCD sum rule approach suggested that fully heavy dibaryons should exist as weakly bound ones \cite{Wang:2022jvk}.
Huang {\it et al.} systematically studied the fully heavy dibaryons, and their results suggested the existence of $\Omega_{ccc}\Omega_{bbb}$ and $\Omega_{ccc}\Omega_{ccc}$ and the absence of $\Omega_{bbb}\Omega_{bbb}$ \cite{Huang:2020bmb}. 
However, many studies within quark models have disfavored the bound fully heavy dibaryons \cite{Richard:2021nvn,Richard:2021jgp,Lu:2022myk,Alcaraz-Pelegrina:2022fsi,Weng:2022ohh,Richard:2020zxb}.
In particular,  Richard {\it et al.} explored the possibility of heavy dibaryons with three charm quarks and three beauty quarks ($bbbccc$) in a potential model, concluding that there is no evidence for a stable state in such a very heavy flavor six-quark system \cite{Richard:2020zxb}.
In addition, Alcaraz-Pelegrina {\it et al.} studied the fully heavy dibaryons with the diffusion Monte Carlo method within the quark model and found that all these states are above the thresholds of the two fully heavy baryons \cite{Alcaraz-Pelegrina:2022fsi}. 
Similarly, Weng {\it et al.} systematically studied the mass spectra of the fully heavy dibaryons in an extended chromomagnetic model and found no stable state below the corresponding baryon-baryon thresholds \cite{Weng:2022ohh}.

In addition to the fully heavy dibaryons,  the stability of the dibaryon system consisting of six quarks or antiquarks has been studied in the context of a string model \cite{Vijande:2011im}, where the existence of the six-quark ($q^6$) states and the three-quark-three-antiquark ($q^3{\bar{q}^3}$) states was discussed, and  the investigations were extended to the ($q^3Q^3$) and ($Q^3\bar{q}^3$) systems. 
Later, Huang {\it et al.} investigated possible $N\Omega$-like dibaryons  $N\Omega_{ccc}$ and $N\Omega_{bbb}$ with quantum numbers $I(J^{P})=1/2(2^{+})$ within the framework of the quark delocalization color screening model \cite{Huang:2019esu}.
Here, both of these states are bound, and the binding energy increases as the quarks of the system become heavier.
In addition, inspired by the LHCb observation of three pentaquark states \cite{LHCb:2019kea}, 
theoretical studies predicted expect the light-quark structures of the $\bar{D}^{(*)}$ and the $\Xi_{cc}^{(*)}$ to be almost identical. Thus,
theoretical work has also focused on the triply heavy dibaryons \cite{Pan:2019skd,Pan:2020xek,Wu:2021kbu,Wang:2019gal,Liu:2018zzu}.

In particular, inspired by the recent discovery of a narrow structure $T^{+}_{cc}$ ($cc\bar{u}\bar{d}$) in the LHCb experiment \cite{LHCb:2021vvq,LHCb:2021auc}, on the basis of the similarity between the $cc\bar{u}\bar{d}$ configuration and the $\Xi^{++}_{cc}$ baryon ($ccu$) \cite{LHCb:2021vvq}, the existence of other open charm multiquarks has been studied via the extended chromomagnetic interaction (CMI) model \cite{Guo:2021mja, Weng:2021hje, Weng:2021ngd}, the molecular state picture \cite{Liu:2011xc,Yang:2019rgw,Meng:2017fwb}, the Born-Oppenheimer approximation \cite{Maiani:2022qze}, and the chiral quark model \cite{Praszalowicz:2022sqx}.

Given this situation, we are interested in further exploring the doubly heavy dibaryon state.
The double heavy dibaryon states ($qqqqQQ$) have been discussed using various approaches, 
including 
the constituent quark model (CQM) \cite{Garcilazo:2020acl,Vijande:2016nzk,Leandri:1997ge},
the quark delocalization color screening model \cite{Xia:2021tof,Huang:2013rla,Huang:2013zva},
the lattice QCD calculation \cite{Junnarkar:2022yak},
the chiral constituent quark model \cite{Carames:2015sya}, 
the nucleon-nucleon interaction model \cite{Julia-Diaz:2004ict,Froemel:2004ea},
the chiral effective field theory (EFT) \cite{Chen:2022iil,Lu:2017dvm,Oka:2013xxa},
the QCD sum rules \cite{Wang:2021pua,Wang:2021qmn},
the relativistic six-quark equations \cite{Gerasyuta:2011zx},
the complex scaling method \cite{Cheng:2022vgy},
and the one-boson-exchange (OBE) model \cite{Ling:2021asz,Chen:2017vai,Meguro:2011nr,Meng:2017udf,Lee:2011rka,Li:2012bt}.
Meanwhile, Richard pointed out that there is no obvious gain with respect to the best threshold made of two baryons in a colour-additive model, but flavour independence dictates that before spin corrections, the threshold ($QQq$) + ($qqq$) is lower than 2 ($Qqq$) \cite{Richard:2016eis}.
In this work, the mass spectrum of these 
doubly heavy dibaryon states will be obtained in the framework of the chromomagnetic model, which is usually used to calculate the mass spectrum of ordinary hadron states or multiquark states \cite{Liu:2019zoy}.
The present paper discusses the mass spectrum of these doubly charmed dibaryon states and briefly discusses the stability of various states. 
It may provide valuable information to further experimental searches for doubly charmed dibaryon states.
With the running of the High-Luminosity LHC \cite{LHCb:2022hl}, the hunt for this new type of dibaryon may become a reality.

The remainder of this paper is organized as follows. In Sec.~\ref{sec2}, we introduce the chromomagentic interaction model. According to different permutation symmetries of identical particles, we construct the $F_{flavor} \otimes \phi_{color} \otimes\chi_{spin}$ wave functions for the $S$-wave $qqqqcc$ dibaryon state. The mass spectrum and decay mode of the $qqqqcc$ hexaquark state with different constituent quarks are discussed in Sec.~\ref{sec4}. Finally, a short summary is given in Sec.~\ref{sec5}.

%%%%%%%%%%%%%%%%%%%%%%%%%%%%%%%%%%%%%%%%%%%%%%%%%%%%%%%%%%%%%%%%%%%%
\section{Hamiltonian in CMI model}\label{sec2}
In the chromomagnetic interaction (CMI) model, 
the mass of the ground hadron state can be described by the effective Hamiltonian \cite{Liu:2021gva,Weng:2022ohh,Weng:2021hje,Weng:2021ngd,An:2021vwi,Guo:2021mja,Guo:2021yws,Weng:2018mmf}
\begin{eqnarray}
H&=&\sum_{i}m_i+H_{\textrm{CI}}+H_{\textrm{CMI}}\nonumber\\
&=&\sum_im_i-\sum_{i<j}A_{ij} \vec\lambda_i\cdot \vec\lambda_j-\sum_{i<j}v_{ij} \vec\lambda_i\cdot \vec\lambda_j \vec\sigma_i\cdot\vec\sigma_j \nonumber\\
&=&-\frac{3}{4}\sum_{i<j}m_{ij}V^{\rm C}_{ij}-\sum_{i<j}v_{ij}V^{\rm CMI}_{ij},
\end{eqnarray}
where $V^{\textrm{C}}_{ij}=\vec\lambda_i\cdot \vec\lambda_j$ and $V^{\textrm{CMI}}_{ij}=\vec\lambda_i\cdot \vec\lambda_j \vec\sigma_i\cdot\vec\sigma_j$
 represent the color and chromomagnetic interaction between the quarks, respectively,
with $\sigma_{i}$ and $\lambda_i$ being the Pauli matrices and Gell-Mann matrices, respectively.
The mass parameter of the quark pair $m_{ij}=1/4(m_{i}+m_{j})+4/3A_{ij}$ contains the effective quark mass $m_{i}$ ($m_{j}$) and the color interaction strength $A_{ij}$.
The $v_{ij}$ is the effective coupling constant between the $i$-th quark and the $j$-th quark, which determines the mass gaps.

To estimate the mass spectrum of the doubly charmed dibaryon states, we extract the effective coupling parameters $m_{ij}$ and $v_{ij}$ from the conventional baryon masses, which are presented in Table \ref{para}. 

\begin{table}[h]
\centering \caption{Relevant coupling parameters for the doubly charmed dibaryon system  \cite{Weng:2018mmf} (in units of MeV).
}\label{para}
\renewcommand\arraystretch{1.3}
\renewcommand\tabcolsep{5pt}
\begin{tabular}{cccccccc}
\bottomrule[1.5pt]
\bottomrule[0.5pt]
%\bottomrule[1.0pt]
%\bottomrule[1.0pt]
Parameter&$m_{nn}$&$m_{ns}$&$m_{ss}$&$m_{nc}$&$m_{sc}$&$m_{cc}$\\
Value&181.2&226.7&262.3&520.0&545.9&792.9\\
\bottomrule[0.7pt]
Parameter&$v_{nn}$&$v_{ns}$&$v_{ss}$&$v_{nc}$&$v_{sc}$&$v_{cc}$\\
Value&19.1&13.3&12.2&3.9&4.4&3.5\\
\bottomrule[0.5pt]
\bottomrule[1.5pt]
\end{tabular}
\end{table}

%%%%%%%%%%%%%%%%%%%%%%%%%%%%%%%%%%%%%%%%%%%%%%%%%%%%%%%%%%%%%%%%%%%%
% \section{The wave functions}
%%%%%%%%%%%%%%%%%%%%%%%%%%%%%%%%%%%%%%%%%%%%%%%%%%%%%%%%%%%%%%%%%
%\label{sec3}

To calculate the mass spectrum of the doubly charmed dibaryon states in the CMI model, we need the information on the total wave function, which is composed of the spatial, flavor, color, and spin wave functions, i.e.,
\begin{equation}
	\Phi_{total}=\Psi_{space}\otimes F_{flavor} \otimes \phi_{color} \otimes\chi_{spin}.
\end{equation}
According to the Pauli principle, this wave function should be completely antisymmetric when identical quarks are exchanged.
Here, we consider only the low-lying $S$-wave dibaryon states, so that their spatial wave functions are symmetric and become trivial. 
Thus, the $F_{flavor} \otimes \phi_{color} \otimes\chi_{spin}$ wave functions of the dibaryon states should be completely antisymmetric when identical quarks are exchanged.
%In Table \ref{flavor}, we list all the possible flavor combinations 
For the doubly charmed dibaryon system,  all the possible flavor combinations are $nnnncc$ ($I=2,1,0$), $nnnscc$ ($I=3/2,1/2$), $nnsscc$ ($I=1,0$), $sssncc$, and $sssscc$.
According to the symmetry properties, we can divide the $qqqqcc$ dibaryon system into three groups, which are presented in Table \ref{flavor}.
Here, we have to construct the corresponding $F_{flavor} \otimes \phi_{color} \otimes\chi_{spin}$ wave functions with the symmetries $\{ 1234\}\{56\}$,  $\{ 123\}4\{56\}$, and  $\{ 12\}\{34\}\{56\}$. 
We use the notation $\{123...\}$, where quarks 1, 2, 3, and so on have antisymmetric properties.

In previous works, there are abundant methods on the wave functions of six-quark systems \cite{Hogaasen:1978xs, Wang:1995kp, Pepin:1998ih, Vijande:2016nzk, Park:2016cmg, Kim:2020rwn}. 
%In previous works, several methods exist to construct the wave functions of six-quark systems \cite{Hogaasen:1978xs, Wang:1995kp, Pepin:1998ih, Vijande:2016nzk, Park:2016cmg, Kim:2020rwn}. 
The light flavor SU(3) symmetry and symmetry breaking have been considered in these works. 
%And the method we used in this paper can be found in Refs.~\cite{Stancu:1999qr, Park:2016mez, An:2020jix, Park:2015nha}."
In addition, the same method we used was applied in Refs.~\cite{Park:2016mez, An:2020jix, Park:2015nha}.

\begin{table}[h]
\centering \caption{All possible flavor combinations for the $qqqqcc$ dibaryon system. Here, $q=n$, $s$ and $n=u$, $d$.
}\label{flavor}
\begin{lrbox}{\tablebox}
\renewcommand\arraystretch{1.5}
\renewcommand\tabcolsep{5pt}
\begin{tabular}{l|lc}
\bottomrule[1.5pt]
\bottomrule[0.5pt]
Symmetry&\multicolumn{2}{c}{Flavor combinations}\\
\bottomrule[0.5pt]
$\{ 1234\}\{56\}$&$nnnncc$ ($I=2,1,0$)&  $sssscc$\\
 $\{ 123\}4\{56\}$& $nnnscc$ ($I=3/2,1/2$)&$sssncc$\\
$\{ 12\}\{34\}\{56\}$&$nnsscc$ ($I=1,0$)\\
\bottomrule[0.5pt]
\bottomrule[1.5pt]
\end{tabular}
\end{lrbox}\scalebox{1.08}{\usebox{\tablebox}}
\end{table}

\subsection{Wave function for $\{ 1234\}\{56\}$ symmetry}
\label{sec:4ip}

\begin{table*}[htp]
\caption{The flavor, color, and spin wave functions represented by the partition and the corresponding Young tableaux. 
Here, the $I/J$ represents the isospin/spin. }\label{wave} \centering
%\begin{minipage}{0.0\linewidth}
%\resizebox{\columnwidth}{!}{
%\renewcommand\arraystretch{2.99}
\begin{tabular}{c|c|c|lll}
\midrule[1.5pt]
\bottomrule[0.5pt]
&$I$/$J$&Partition&\multicolumn{3}{c}{The Young Tableau}\\
%The color-spin wave function\\
\hline
\multirow{6}*{\makecell[c]{Flavor\\\\ part}}&\multirow{1}*{$I=2$}&[4]&
$F^{a}_{1}=(\begin{tabular}{|c|c|c|c|c|}
\hline
1&2&3&4  \\
\cline{1-4}
\end{tabular}_{\begin{tabular}{|c|} \multicolumn{1}{c}{$I$}\end{tabular}}
\begin{tabular}{|c|c|}
\hline
5&6  \\
\cline{1-2}
\end{tabular}
)_{\begin{tabular}{|c|} \multicolumn{1}{c}{$F$}\end{tabular}}$\\ 
&\multirow{1}*{$I=1$}&[3,1]&
$F^{a}_{2}=(\begin{tabular}{|c|c|c|c|c|}
\hline
1&2&3  \\
\cline{1-3}
4  \\
\cline{1-1}
\end{tabular}_{\begin{tabular}{|c|} \multicolumn{1}{c}{$I$}\end{tabular}}
\begin{tabular}{|c|c|}
\hline
5&6  \\
\cline{1-2}
\end{tabular}
)_{\begin{tabular}{|c|} \multicolumn{1}{c}{$F$}\end{tabular}}$&
$F^{a}_{3}=(\begin{tabular}{|c|c|c|c|c|}
\hline
1&2&4  \\
\cline{1-3}
3  \\
\cline{1-1}
\end{tabular}_{\begin{tabular}{|c|} \multicolumn{1}{c}{$I$}\end{tabular}}
\begin{tabular}{|c|c|}
\hline
5&6  \\
\cline{1-2}
\end{tabular}
)_{\begin{tabular}{|c|} \multicolumn{1}{c}{$F$}\end{tabular}}$&
$F^{a}_{4}=(\begin{tabular}{|c|c|c|c|c|}
\hline
1&3&4  \\
\cline{1-3}
2  \\
\cline{1-1}
\end{tabular}_{\begin{tabular}{|c|} \multicolumn{1}{c}{$I$}\end{tabular}}
\begin{tabular}{|c|c|}
\hline
5&6  \\
\cline{1-2}
\end{tabular})_{\begin{tabular}{|c|} \multicolumn{1}{c}{$F$}\end{tabular}}$
\\ 
&\multirow{1}*{$I=0$}&[2,2]&
$F^{a}_{5}=(\begin{tabular}{|c|c|c|c|c|}
\hline
1&2  \\
\cline{1-2}
3&4  \\
\cline{1-2}
\end{tabular}_{\begin{tabular}{|c|} \multicolumn{1}{c}{$I$}\end{tabular}}
\begin{tabular}{|c|c|}
\hline
5&6  \\
\cline{1-2}
\end{tabular})_{\begin{tabular}{|c|} \multicolumn{1}{c}{$F$}\end{tabular}}$
&
$F^{a}_{6}=(\begin{tabular}{|c|c|c|c|c|}
\hline
1&3  \\
\cline{1-2}
2&4  \\
\cline{1-2}
\end{tabular}_{\begin{tabular}{|c|} \multicolumn{1}{c}{$I$}\end{tabular}}
\begin{tabular}{|c|c|}
\hline
5&6  \\
\cline{1-2}
\end{tabular})_{\begin{tabular}{|c|} \multicolumn{1}{c}{$F$}\end{tabular}}$\\
%The color-spin wave function\\
\hline
\multicolumn{2}{c|}{Color part}&[2,2,2]&\multicolumn{3}{l}{
$\phi^{a}_{1}=\begin{tabular}{|c|c|c|c|c|}
\hline
1&2  \\
\cline{1-2}
3&5  \\
\cline{1-2}
4&6  \\
\cline{1-2}
\end{tabular}_{\begin{tabular}{|c|} \multicolumn{1}{c}{$C$}\end{tabular}}$
$\phi^{a}_{2}=\begin{tabular}{|c|c|c|c|c|}
\hline
1&2  \\
\cline{1-2}
3&4  \\
\cline{1-2}
5&6  \\
\cline{1-2}
\end{tabular}_{\begin{tabular}{|c|} \multicolumn{1}{c}{$C$}\end{tabular}}$
$\phi^{a}_{3}=\begin{tabular}{|c|c|c|c|c|}
\hline
1&3  \\
\cline{1-2}
2&4  \\
\cline{1-2}
5&6  \\
\cline{1-2}
\end{tabular}_{\begin{tabular}{|c|} \multicolumn{1}{c}{$C$}\end{tabular}}$
$\phi^{a}_{4}=\begin{tabular}{|c|c|c|c|c|}
\hline
1&3  \\
\cline{1-2}
2&5  \\
\cline{1-2}
4&6  \\
\cline{1-2}
\end{tabular}_{\begin{tabular}{|c|} \multicolumn{1}{c}{$C$}\end{tabular}}$
$\phi^{a}_{5}=\begin{tabular}{|c|c|c|c|c|}
\hline
1&4  \\
\cline{1-2}
2&5  \\
\cline{1-2}
3&6  \\
\cline{1-2}
\end{tabular}_{\begin{tabular}{|c|} \multicolumn{1}{c}{$C$}\end{tabular}}$
}\\
\hline
\multirow{10}*{\makecell[c]{Spin\\\\ part}}&\multirow{1}*{$J=3$}&[6]&
$\chi^{a}_{1}=\begin{tabular}{|c|c|c|c|c|c|}
\hline
1&2&3&4&5&6  \\
\cline{1-6}
\end{tabular}_{\begin{tabular}{|c|} \multicolumn{1}{c}{$S$}\end{tabular}}$\\
&\multirow{1}*{$J=2$}&[5,1]&
\multicolumn{3}{l}{$\chi^{a}_{2}=\begin{tabular}{|c|c|c|c|c|c|}
\hline
1&2&3&4&5  \\
\cline{1-5}
6  \\
\cline{1-1}
\end{tabular}_{\begin{tabular}{|c|} \multicolumn{1}{c}{$S$}\end{tabular}}$
$\chi^{a}_{3}=\begin{tabular}{|c|c|c|c|c|c|}
\hline
1&2&3&4&6  \\
\cline{1-5}
5  \\
\cline{1-1}
\end{tabular}_{\begin{tabular}{|c|} \multicolumn{1}{c}{$S$}\end{tabular}}$
$\chi^{a}_{4}=\begin{tabular}{|c|c|c|c|c|c|}
\hline
1&2&3&5&6  \\
\cline{1-5}
4  \\
\cline{1-1}
\end{tabular}_{\begin{tabular}{|c|} \multicolumn{1}{c}{$S$}\end{tabular}}$
$\chi^{a}_{5}=\begin{tabular}{|c|c|c|c|c|c|}
\hline
1&2&4&5&6  \\
\cline{1-5}
3  \\
\cline{1-1}
\end{tabular}_{\begin{tabular}{|c|} \multicolumn{1}{c}{$S$}\end{tabular}}$
$\chi^{a}_{6}=\begin{tabular}{|c|c|c|c|c|c|}
\hline
1&3&4&5&6  \\
\cline{1-5}
2  \\
\cline{1-1}
\end{tabular}_{\begin{tabular}{|c|} \multicolumn{1}{c}{$S$}\end{tabular}}$
}\\
&\multirow{4}*{$J=1$}&\multirow{4}*{[4,2]}&
\multicolumn{3}{l}{$\chi^{a}_{7}=\begin{tabular}{|c|c|c|c|c|c|}
\hline
1&2&3&4  \\
\cline{1-4}
5&6  \\
\cline{1-2}
\end{tabular}_{\begin{tabular}{|c|} \multicolumn{1}{c}{$S$}\end{tabular}}$
\quad
$\chi^{a}_{8}=\begin{tabular}{|c|c|c|c|c|c|}
\hline
1&2&3&5  \\
\cline{1-4}
4&6  \\
\cline{1-2}
\end{tabular}_{\begin{tabular}{|c|} \multicolumn{1}{c}{$S$}\end{tabular}}$
\quad
$\chi^{a}_{9}=\begin{tabular}{|c|c|c|c|c|c|}
\hline
1&2&3&6  \\
\cline{1-4}
4&5  \\
\cline{1-2}
\end{tabular}_{\begin{tabular}{|c|} \multicolumn{1}{c}{$S$}\end{tabular}}$
\quad
$\chi^{a}_{10}=\begin{tabular}{|c|c|c|c|c|c|}
\hline
1&2&4&5  \\
\cline{1-4}
3&6  \\
\cline{1-2}
\end{tabular}_{\begin{tabular}{|c|} \multicolumn{1}{c}{$S$}\end{tabular}}$
\quad
$\chi^{a}_{11}=\begin{tabular}{|c|c|c|c|c|c|}
\hline
1&2&4&6  \\
\cline{1-4}
3&5  \\
\cline{1-2}
\end{tabular}_{\begin{tabular}{|c|} \multicolumn{1}{c}{$S$}\end{tabular}}$
}\\
&&&
\multicolumn{3}{l}{$\chi^{a}_{12}=\begin{tabular}{|c|c|c|c|c|c|}
\hline
1&2&5&6  \\
\cline{1-4}
3&4  \\
\cline{1-2}
\end{tabular}_{\begin{tabular}{|c|} \multicolumn{1}{c}{$S$}\end{tabular}}$
$\chi^{a}_{13}=\begin{tabular}{|c|c|c|c|c|c|}
\hline
1&3&4&5  \\
\cline{1-4}
2&6  \\
\cline{1-2}
\end{tabular}_{\begin{tabular}{|c|} \multicolumn{1}{c}{$S$}\end{tabular}}$
$\chi^{a}_{14}=\begin{tabular}{|c|c|c|c|c|c|}
\hline
1&3&4&6  \\
\cline{1-4}
2&5  \\
\cline{1-2}
\end{tabular}_{\begin{tabular}{|c|} \multicolumn{1}{c}{$S$}\end{tabular}}$
$\chi^{a}_{15}=\begin{tabular}{|c|c|c|c|c|c|}
\hline
1&3&5&6  \\
\cline{1-4}
2&4  \\
\cline{1-2}
\end{tabular}_{\begin{tabular}{|c|} \multicolumn{1}{c}{$S$}\end{tabular}}$
}\\
&\multirow{1}*{$J=0$}&\multirow{1}*{[3,3]}&
\multicolumn{3}{l}{$\chi^{a}_{16}=\begin{tabular}{|c|c|c|c|c|c|}
\hline
1&2&3  \\
\cline{1-3}
4&5&6  \\
\cline{1-3}
\end{tabular}_{\begin{tabular}{|c|} \multicolumn{1}{c}{$S$}\end{tabular}}$
\quad
$\chi^{a}_{17}=\begin{tabular}{|c|c|c|c|c|c|}
\hline
1&2&4  \\
\cline{1-3}
3&5&6  \\
\cline{1-3}
\end{tabular}_{\begin{tabular}{|c|} \multicolumn{1}{c}{$S$}\end{tabular}}$
\quad
$\chi^{a}_{18}=\begin{tabular}{|c|c|c|c|c|c|}
\hline
1&2&5  \\
\cline{1-3}
3&4&6  \\
\cline{1-3}
\end{tabular}_{\begin{tabular}{|c|} \multicolumn{1}{c}{$S$}\end{tabular}}$
\quad
$\chi^{a}_{19}=\begin{tabular}{|c|c|c|c|c|c|}
\hline
1&3&4  \\
\cline{1-3}
2&5&6  \\
\cline{1-3}
\end{tabular}_{\begin{tabular}{|c|} \multicolumn{1}{c}{$S$}\end{tabular}}$
\quad
$\chi^{a}_{20}=\begin{tabular}{|c|c|c|c|c|c|}
\hline
1&3&5  \\
\cline{1-3}
2&4&6  \\
\cline{1-3}
\end{tabular}_{\begin{tabular}{|c|} \multicolumn{1}{c}{$S$}\end{tabular}}$
}\\
\bottomrule[0.5pt]
\midrule[1.5pt]
\end{tabular}
\end{table*}

First, we construct the wave function of the $nnnncc$ ($I=2,1,0$) and $sssscc$ dibaryon states.
Because the $nnnncc$ ($I=2$) dibaryon states and $sssscc$ dibaryon states have exactly the same symmetry requirements, we only need to concentrate on the $nnnncc$ ($I=2,1,0$) dibaryon states.
Here, we introduce the Young tableau to represent the irreducible base of the permutation group $S_n$. 
It can help us to attach certain symmetry properties to the total wave function of the $nnnncc$ dibaryon states.
For the $nnnncc$ dibaryon states with isospin $I=2,1,$ and $0$, the flavored states of four $n$-quarks can be represented by partitions [4], [3,1], and [2,2], respectively, and specific Young tableaux are given in Table \ref{wave}.
Owing to the requirement of color confinement, the color part must be a singlet in the SU(3) symmetry. 
Thus, only the partition [2,2,2] is satisfied with the color singlet.
The spin part can be represented by the partitions [6], [5,1], [4,2], and [3,3] for total spin of $J=3,2,1,$ and $0$, respectively.
Accordingly, we present the specific Young tableaux in Table \ref{wave}.

Once we have obtained the wave functions represented by the Young tableaux, the next step is to adopt a suitable coupling method to combine these wave functions in different spaces into fully antisymmetric wave functions. Thus, we need to introduce the corresponding Clebsch-Gordan (CG) coefficients of the permutation group $S_{n}$, similar to the coupling method used in Refs. \cite{Stancu:1999qr, Park:2016mez,Park:2015nha,An:2020jix}
\begin{equation}
	|[f]Y \rangle=\sum_{Y',Y''}S([[f']Y'[f'']Y'']|[f]Y)|[f']Y' \rangle|[f'']Y'' \rangle.
\end{equation}
Here, $[f]$, $[f']$, or $[f'']$ denotes an irreducible representation of $S_{n}$, $Y$ denotes a Young tableau and $|[f]Y \rangle$ is a basis vector of $S_{n}$. $S([[f']Y'[f'']Y'']|[f]Y)|[f']Y' \rangle$ are CG coefficients. As mentioned above, the symmetry of the color wave function is fixed as $[2,2,2]$. Thus we adopt a coupling method whereby one can first combine the color-SU(3) singlet $[2,2,2]$ with spin-SU(2) representation $[f]_{S}$ into an SU(6) $[f]_{CS}$ and then combine it with isospin-SU(2) representations $[f]_{I}$ to obtain SU(12) representations $[f]_{CSI}$.
Thus, we can calculate the 6-quark wave function with the desired permutation symmetry $[f]$ as a linear combination of $F_{flavor} \otimes \phi_{color} \otimes\chi_{spin}$.

\subsection{Wave function for $\{123\}4\{56\}$ symmetry}
\label{sec:3ip}

Next, we construct the wave functions of the $nnnscc$ ($I=3/2,1/2$) and $sssncc$ states.
Because the $nnnscc$ ($I=3/2$) states and $sssncc$  states have exactly the same symmetry requirements, 
we only need to concentrate on the $nnnscc$ ($I=3/2,1/2$) states.
The $s$ quark is far heavier than the $u$ and $d$ quarks, leading to an SU(3) flavor symmetry breaking effect.
We therefore include this effect by distinguishing the $s$ quark from the $u$ and $d$ quarks, while still keeping the SU(2) isospin symmetry in our calculation.
Accordingly, it is convenient to use the $|[(nn)n][s(cc)]\rangle$ basis to construct the total wave functions of these states.
We list all the possible flavours, color singlets, and spin wave functions in Table \ref{base}.

In Table \ref{base}, S(A) means fully symmetric (antisymmetric), 
and MS(MA) means that the first two quarks $nn$ or the last two quarks $cc$ are symmetric (antisymmetric) in $nnnscc$ states. 
In the color part, we use the notation $|[(n_{1}n_{2})^{\rm color1}n_{3}]^{\rm color2}[s_{4}(c_{5}c_{6})^{\rm color3}]^{\rm color4}\rangle$ to describe the color singlet wave functions, where
color1, color2, color3, and color4 stand for the color representations of $n_{1}n_{2}$, $n_{1}n_{2}n_{3}$, $c_{5}c_{6}$, and $s_{4}c_{5}c_{6}$, respectively.
Similarly, we use the notation $|[(n_{1}n_{2})_{\rm spin1}n_{3}]_{\rm spin2}[s_{4}(c_{5}c_{6})_{\rm spin3}]_{\rm spin4}\rangle_{\rm spin5}$ to describe the spin wave functions, where
spin1, spin2, spin3, spin4, and spin5 represent the spins of $n_{1}n_{2}$, $n_{1}n_{2}n_{3}$, $c_{5}c_{6}$, and $s_{4}c_{5}c_{6}$ and total spin, respectively.

\begin{table*}[h]
\caption{All the possible flavors, colors, and spin wave functions for the $nnnscc$ and $nnsscc$ states. $F\otimes \phi \otimes \chi$ represents the combination of the wave functions in different spaces.
Here, the $I/J$ represents the isospin/spin.
}\label{base}
\begin{lrbox}{\tablebox}
\renewcommand\arraystretch{1.5}
\renewcommand\tabcolsep{2.85pt}
\begin{tabular}{c|c|cccccc}
\toprule[1.0pt]
\toprule[0.5pt]
\multicolumn{8}{c}{$nnnscc$ states}\\
\toprule[0.5pt]
\multirow{2}*{Flavor}&\multicolumn{1}{c|}{$I=\frac{3}{2}$}&\multicolumn{2}{r}{$F^{S,S}=|[(nn)^{I=1}n]^{I=3/2}[s(cc)]\rangle$}\\
&\multicolumn{1}{c|}{$I=\frac{1}{2}$}&\multicolumn{2}{r}{$F^{MS,S}=|[(nn)^{I=1}n]^{I=1/2}[s(cc)]\rangle$}&
\multicolumn{2}{r}{$F^{MA,S}=|[(nn)^{I=0}n]^{I=1/2}[s(cc)]\rangle$}\\
\toprule[0.5pt]
\multirow{2}*{Color}&\multicolumn{1}{c}{}&
\multicolumn{2}{r}{$\phi^{MA,MA}=|[(nn)^{\bar{3}}n]^{8}[s(cc)^{\bar{3}}]^{8}\rangle$}&
\multicolumn{2}{r}{$\phi^{MA,MS}=|[(nn)^{\bar{3}}n]^{8}[s(cc)^{6}]^{8}\rangle$}&
\multicolumn{2}{r}{$\phi^{A,A}=|[(nn)^{\bar{3}}n]^{1}[s(cc)^{\bar{3}}]^{1}\rangle$}\\
&\multicolumn{1}{c}{}&\multicolumn{2}{r}{$\phi^{MS,MA}=|[(nn)^{6}n]^{8}[s(cc)^{\bar{3}}]^{8}\rangle$}&
\multicolumn{2}{r}{$\phi^{MS,MS}=|[(nn)^{6}n]^{8}[s(cc)^{6}]^{8}\rangle$}\\
\toprule[0.5pt]
\multirow{8}*{Spin}&\multicolumn{1}{c|}{$J=3$}&
\multicolumn{2}{r}{$\chi_{1}^{S,S}=|[(nn)_{1}n]_{3/2}[s(cc)_{1}]_{3/2}\rangle_{3}$}\\
&\multirow{2}*{$J=2$}
&\multicolumn{2}{r}{$\chi_{2}^{S,S}=|[(nn)_{1}n]_{3/2}[s(cc)_{1}]_{3/2}\rangle_{2}$}
&\multicolumn{2}{r}{$\chi_{3}^{S,MS}=|[(nn)_{1}n]_{3/2}[s(cc)_{1}]_{1/2}\rangle_{2}$}\\
&&\multicolumn{2}{r}{$\chi_{4}^{MS,S}=|[(nn)_{1}n]_{1/2}[s(cc)_{1}]_{3/2}\rangle_{2}$}&
\multicolumn{2}{r}{$\chi_{5}^{MS,A}=|[(nn)_{1}n]_{3/2}[s(cc)_{0}]_{3/2}\rangle_{2}$}&
\multicolumn{2}{r}{$\chi_{6}^{MA,S}=|[(nn)_{0}n]_{1/2}[s(cc)_{1}]_{3/2}\rangle_{2}$}\\
&\multirow{3}*{$J=1$}&
\multicolumn{2}{r}{$\chi_{7}^{S,S}=|[(nn)_{1}n]_{3/2}[s(cc)_{1}]_{3/2}\rangle_{1}$}&
\multicolumn{2}{r}{$\chi_{8}^{S,MS}=|[(nn)_{1}n]_{3/2}[s(cc)_{1}]_{1/2}\rangle_{1}$}&
\multicolumn{2}{r}{$\chi_{9}^{MS,S}=|[(nn)_{1}n]_{1/2}[s(cc)_{1}]_{3/2}\rangle_{1}$}\\
&&\multicolumn{2}{r}{\quad$\chi_{10}^{MS,MS}=|[(nn)_{1}n]_{1/2}[s(cc)_{1}]_{1/2}\rangle_{1}$}&
\multicolumn{2}{r}{$\chi_{11}^{S,MA}=|[(nn)_{1}n]_{3/2}[s(cc)_{0}]_{1/2}\rangle_{1}$}&
\multicolumn{2}{r}{$\chi_{12}^{MA,S}=|[(nn)_{0}n]_{1/2}[s(cc)_{1}]_{3/2}\rangle_{1}$}\\
&&\multicolumn{2}{r}{$\chi_{13}^{MS,MA}=|[(nn)_{1}n]_{1/2}[s(cc)_{0}]_{1/2}\rangle_{1}$}&
\multicolumn{2}{r}{\quad\quad$\chi_{14}^{MA,MS}=|[(nn)_{0}n]_{1/2}[s(cc)_{1}]_{1/2}\rangle_{1}$}&
\multicolumn{2}{r}{$\chi_{15}^{MA,MA}=|[(nn)_{0}n]_{1/2}[s(cc)_{0}]_{1/2}\rangle_{1}$}\\
&\multirow{2}*{$J=0$}&\multicolumn{2}{r}{$\chi_{16}^{S,S}=|[(nn)_{1}n]_{3/2}[s(cc)_{1}]_{3/2}\rangle_{0}$}&
\multicolumn{2}{r}{$\chi_{17}^{MS,MS}=|[(nn)_{1}n]_{1/2}[s(cc)_{1}]_{1/2}\rangle_{0}$}&
\multicolumn{2}{r}{$\chi_{18}^{MS,MA}=|[(nn)_{1}n]_{1/2}[s(cc)_{0}]_{1/2}\rangle_{0}$}\\
&&\multicolumn{2}{r}{$\chi_{19}^{MA,MS}=|[(nn)_{0}n]_{1/2}[s(cc)_{1}]_{1/2}\rangle_{1}$}&
\multicolumn{2}{r}{$\chi_{20}^{MA,MA}=|[(nn)_{0}n]_{1/2}[s(cc)_{0}]_{1/2}\rangle_{0}$}\\
\toprule[0.5pt]
\multirow{15}*{\makecell[c]{$F$\\ $\otimes$\\ $\phi$\\ $\otimes$\\$\chi$}}&\multirow{3}*{$I=\frac{3}{2}$}&
\multicolumn{3}{l}{\quad$F^{S,S}\otimes\phi^{A,A}\otimes\chi^{S,S}$;\quad\quad$F^{S,S}\otimes\phi^{A,A}\otimes\chi^{S,MS}$;}&
\multicolumn{3}{r}{$\frac{1}{\sqrt{2}}[F^{S,S}\otimes(\phi^{MS,MA}\otimes\chi^{MA,MS}-\phi^{MA,MA}\otimes\chi^{MS,S})]$}\\
&&\multicolumn{3}{l}{
$\frac{1}{\sqrt{2}}[F^{S,S}\otimes(\phi^{MS,MA}\otimes\chi^{MA,S}-\phi^{MA,MA}\otimes\chi^{MS,S})]$;}&
\multicolumn{3}{r}{$\frac{1}{\sqrt{2}}[F^{S,S}\otimes(\phi^{MS,MS}\otimes\chi^{MA,MA}-\phi^{MA,MS}\otimes\chi^{MS,MA})]$}\\
&&\multicolumn{3}{l}{$\frac{1}{\sqrt{2}}[F^{S,S}\otimes(\phi^{MS,MA}\otimes\chi^{MA,S}-\phi^{MA,MA}\otimes\chi^{MS,MS})]$;}&
\multicolumn{3}{r}{
$\frac{1}{\sqrt{2}}[F^{S,S}\otimes(\phi^{MS,MA}\otimes\chi^{MA,MS}-\phi^{MA,MA}\otimes\chi^{MS,MS})]$}\\
\Xcline{2-8}{0.5pt}
&\multirow{12}*{$I=\frac{1}{2}$}
&\multicolumn{3}{l}{
$\frac{1}{\sqrt{2}}[(F^{MS,S}\otimes\phi^{MA,MA}-F^{MA,S}\otimes\phi^{MS,MA})\otimes\chi^{S,S})]$;}&
\multicolumn{3}{r}{
$\frac{1}{\sqrt{2}}(F^{MS,S}\otimes\phi^{MA,MA}\otimes\chi^{S,S}-F^{MA,S}\otimes\phi^{MS,MS}\otimes\chi^{S,MA})$}\\
&&\multicolumn{3}{r}{
$\frac{1}{\sqrt{2}}[(F^{MS,S}\otimes\phi^{MA,MA}-F^{MA,S}\otimes\phi^{MS,MA})\otimes\chi^{S,MS})]$;}&
\multicolumn{3}{r}{
$\frac{1}{\sqrt{2}}(F^{MS,S}\otimes\phi^{MA,MS}\otimes\chi^{S,MA}-F^{MA,S}\otimes\phi^{MS,MA}\otimes\chi^{S,MS})$}\\
&&\multicolumn{3}{l}{
$\frac{1}{\sqrt{2}}[(F^{MS,S}\otimes\phi^{MA,MS}-F^{MA,S}\otimes\phi^{MA,MS})\otimes\chi^{S,MA})]$;}&
\multicolumn{3}{r}{
$\frac{1}{\sqrt{2}}(F^{MS,S}\otimes\phi^{MA,MA}\otimes\chi^{S,MS}-F^{MA,S}\otimes\phi^{MS,MS}\otimes\chi^{S,MA})$}\\
&&\multicolumn{6}{r}{
$1/2[(F^{MS,S}\otimes\phi^{MS,MS}-F^{MA,S}\otimes\phi^{MA,MS})\otimes\chi^{MA,MA})+
(F^{MS,S}\otimes\phi^{MA,MS}+F^{MA,S}\otimes\phi^{MS,MS})\otimes\chi^{MS,MA})]$}\\
&&\multicolumn{6}{r}{
$1/2[(F^{MS,S}\otimes\phi^{MS,MA}-F^{MA,S}\otimes\phi^{MA,MA})\otimes\chi^{MA,MS})+
(F^{MS,S}\otimes\phi^{MA,MS}+F^{MA,S}\otimes\phi^{MS,MS})\otimes\chi^{MS,MA})]$}\\
&&\multicolumn{6}{r}{
$1/2[(F^{MS,S}\otimes\phi^{MS,MA}-F^{MA,S}\otimes\phi^{MA,MA})\otimes\chi^{MA,MS})+
(F^{MS,S}\otimes\phi^{MA,MA}+F^{MA,S}\otimes\phi^{MS,MA})\otimes\chi^{MS,MS})]$}\\
&&\multicolumn{6}{r}{
$1/2[(F^{MS,S}\otimes\phi^{MS,MS}-F^{MA,S}\otimes\phi^{MA,MS})\otimes\chi^{MA,MA})+
(F^{MS,S}\otimes\phi^{MA,MA}+F^{MA,S}\otimes\phi^{MS,MA})\otimes\chi^{MS,MS})]$}\\
&&\multicolumn{6}{r}{
$1/2[(F^{MS,S}\otimes\phi^{MS,MA}-F^{MA,S}\otimes\phi^{MA,MA})\otimes\chi^{MA,S})+
(F^{MS,S}\otimes\phi^{MA,MS}+F^{MA,S}\otimes\phi^{MS,MS})\otimes\chi^{MS,MA})]$}\\
&&\multicolumn{6}{r}{
$1/2[(F^{MS,S}\otimes\phi^{MS,MA}-F^{MA,S}\otimes\phi^{MA,MA})\otimes\chi^{MA,S})+
(F^{MS,S}\otimes\phi^{MA,MA}+F^{MA,S}\otimes\phi^{MS,MA})\otimes\chi^{MS,MS})]$}\\
&&\multicolumn{6}{r}{
$1/2[(F^{MS,S}\otimes\phi^{MS,MA}-F^{MA,S}\otimes\phi^{MA,MA})\otimes\chi^{MA,MS})+
(F^{MS,S}\otimes\phi^{MA,MA}+F^{MA,S}\otimes\phi^{MS,MA})\otimes\chi^{MS,S})]$}\\
&&\multicolumn{6}{r}{
$1/2[(F^{MS,S}\otimes\phi^{MS,MS}-F^{MA,S}\otimes\phi^{MA,MS})\otimes\chi^{MA,MA})+
(F^{MS,S}\otimes\phi^{MA,MA}+F^{MA,S}\otimes\phi^{MS,MA})\otimes\chi^{MS,S})]$}\\
&&\multicolumn{6}{r}{
$1/2[(F^{MS,S}\otimes\phi^{MS,MA}-F^{MA,S}\otimes\phi^{MA,MA})\otimes\chi^{MA,S})+
(F^{MS,S}\otimes\phi^{MA,MA}+F^{MA,S}\otimes\phi^{MS,MA})\otimes\chi^{MS,S})]$}\\
\toprule[0.5pt]
\multicolumn{8}{c}{$nnsscc$ states}\\
\toprule[0.5pt]
\multirow{1}*{Flavor}&\multicolumn{1}{c|}{$I=1$}&\multicolumn{2}{r|}{$F^{S,S,S}=|(nn)^{I=1}(ss)(cc)\rangle$}&&
\multicolumn{1}{r|}{$I=0$}&\multicolumn{2}{r}{$F^{A,S,S}=|(nn)^{I=0}(ss)(cc)\rangle$}\\
\toprule[0.5pt]
\multirow{2}*{Color}&\multicolumn{1}{c}{}&
\multicolumn{2}{r}{$\phi^{A,A,A}=|(nn)^{\bar{3}}(ss)^{\bar{3}}(cc)^{\bar{3}}\rangle$}&
\multicolumn{2}{r}{$\phi^{A,S,A}=|(nn)^{\bar{3}}(ss)^{6}(cc)^{\bar{3}}\rangle$}&
\multicolumn{2}{r}{$\phi^{A,A,S}=|(nn)^{\bar{3}}(ss)^{\bar{3}}(cc)^{6}\rangle$}\\
&\multicolumn{1}{c}{}
&\multicolumn{2}{r}{$\phi^{S,A,A}=|(nn)^{6}(ss)^{\bar{3}}(cc)^{\bar{3}}\rangle$}
&\multicolumn{2}{r}{$\phi^{S,S,S}=|(nn)^{6}(ss)^{6}(cc)^{6}\rangle$}\\
\toprule[0.5pt]
\multirow{8}*{Spin}&\multicolumn{1}{c|}{$J=3$}
&\multicolumn{2}{r}{$\chi_{1}^{S,S,S}=|[(nn)_{1}(ss)_{1}]_{2}(cc)_{1}\rangle_{3}$}\\
&\multirow{2}*{$J=2$}&\multicolumn{2}{r}{$\chi_{2}^{S,S,S}=|[(nn)_{1}(ss)_{1}]_{2}(cc)_{1}\rangle_{2}$}
&\multicolumn{2}{r}{$\chi_{3}^{S,S,S}=|[(nn)_{1}(ss)_{1}]_{1}(cc)_{1}\rangle_{2}$}\\
&&\multicolumn{2}{r}{$\chi_{4}^{S,A,S}=|[(nn)_{1}(ss)_{0}]_{1}(cc)_{1}\rangle_{2}$}
&\multicolumn{2}{r}{$\chi_{5}^{S,S,A}=|[(nn)_{1}(ss)_{1}]_{2}(cc)_{0}\rangle_{2}$}
&\multicolumn{2}{r}{$\chi_{6}^{A,S,S}=|[(nn)_{0}(ss)_{1}]_{1}(cc)_{1}\rangle_{2}$}\\
&\multirow{3}*{$J=1$}
&\multicolumn{2}{r}{$\chi_{7}^{S,S,S}=|[(nn)_{1}(ss)_{1}]_{2}(cc)_{1}\rangle_{1}$}
&\multicolumn{2}{r}{$\chi_{8}^{S,S,S}=|[(nn)_{1}(ss)_{1}]_{1}(cc)_{1}\rangle_{1}$}
&\multicolumn{2}{r}{$\chi_{9}^{S,S,S}=|[(nn)_{1}(ss)_{1}]_{0}(cc)_{1}\rangle_{1}$}\\
&&\multicolumn{2}{r}{$\chi_{11}^{S,A,S}=|[(nn)_{1}(ss)_{0}]_{1}(cc)_{1}\rangle_{1}$}
&\multicolumn{2}{r}{$\chi_{12}^{A,S,S}=|[(nn)_{0}(ss)_{1}]_{1}(cc)_{1}\rangle_{1}$}
&\multicolumn{2}{r}{$\chi_{13}^{S,A,A}=|[(nn)_{1}(ss)_{0}]_{1}(cc)_{0}\rangle_{1}$}\\
&&\multicolumn{2}{r}{$\chi_{13}^{S,A,A}=|[(nn)_{1}(ss)_{0}]_{1}(cc)_{0}\rangle_{1}$}
&\multicolumn{2}{r}{$\chi_{14}^{A,S,A}=|[(nn)_{0}(ss)_{1}]_{1}(cc)_{0}\rangle_{1}$}
&\multicolumn{2}{r}{$\chi_{15}^{A,A,S}=|[(nn)_{0}(ss)_{0}]_{0}(cc)_{1}\rangle_{1}$}\\
&\multirow{2}*{$J=0$}&
\multicolumn{2}{r}{$\chi_{16}^{S,S,S}=|[(nn)_{1}(ss)_{1}]_{1}(cc)_{1}\rangle_{0}$}
&\multicolumn{2}{r}{$\chi_{17}^{S,A,S}=|[(nn)_{1}(ss)_{0}]_{1}(cc)_{1}\rangle_{0}$}
&\multicolumn{2}{r}{$\chi_{18}^{S,S,S}=|[(nn)_{1}(ss)_{1}]_{0}(cc)_{0}\rangle_{0}$}\\
&&\multicolumn{2}{r}{$\chi_{19}^{A,S,S}=|[(nn)_{0}(ss)_{1}]_{1}(cc)_{1}\rangle_{0}$}&
\multicolumn{2}{r}{$\chi_{20}^{A,A,A}=|[(nn)_{0}(ss)_{0}]_{0}(cc)_{0}\rangle_{0}$}\\
\toprule[0.5pt]
\multirow{3}*{\makecell[c]{\quad$F$\\ $\otimes$ $\phi$\\ $\otimes$ $\chi$}}&\multirow{2}*{$I=1$}&
\multicolumn{6}{c}{$F^{S,S,S}\otimes\phi^{A,A,A}\otimes\chi^{S,S,S}$;$F^{S,S,S}\otimes\phi^{A,A,S}\otimes\chi^{S,S,A}$;
$F^{S,S,S}\otimes\phi^{A,S,A}\otimes\chi^{S,A,S}$;$F^{S,S,S}\otimes\phi^{S,A,A}\otimes\chi^{A,S,S}$}\\
\Xcline{6-8}{0.5pt}
&&\multicolumn{3}{l|}{\quad\quad\quad\quad$F^{S,S,S}\otimes\phi^{S,S,S}\otimes\chi^{A,A,A}$}&\multicolumn{3}{r}{
$F^{A,S,S}\otimes\phi^{S,S,S}\otimes\chi^{S,A,A}$\quad\quad\quad\quad}\\
\Xcline{2-5}{0.5pt}
&\multirow{1}*{$I=0$}&
\multicolumn{6}{c}{$F^{A,S,S}\otimes\phi^{A,A,A}\otimes\chi^{A,S,S}$;$F^{A,S,S}\otimes\phi^{A,A,S}\otimes\chi^{A,S,A}$;
$F^{A,S,S}\otimes\phi^{A,S,A}\otimes\chi^{A,A,S}$;$F^{A,S,S}\otimes\phi^{S,A,A}\otimes\chi^{S,S,S}$}\\
\toprule[0.5pt]
\toprule[1.0pt]
\end{tabular}
\end{lrbox}\scalebox{0.95}{\usebox{\tablebox}}
\end{table*}

\subsection{Wave function for $\{ 12\}\{ 34\}\{56\}$ symmetry}
\label{sec:2ip}

Finally, we construct the wave function of the $nnsscc$ ($I = 1,0$) dibaryon states.
It is easy to see that there are three pairs of identical particles $nn$, $ss$, and $cc$. 
For these states,
it is convenient to use the $|(nn)(ss)(cc)\rangle$ basis to construct the $nnsscc$ ($I=1,0$) dibaryon states.
We list the corresponding flavours, color singlets, and spin wave functions in Table \ref{base}.
In color part, we use the notation $|(nn)^{\rm color1}(ss)^{\rm color2}(cc)^{\rm color3}\rangle$ to describe the color-singlet wave functions, where
color1, color2, and color3 stand for the color representations of $n_{1}n_{2}$, $s_{3}s_{4}$, and $c_{5}c_{6}$, respectively.
Similarly, we use the notation $|[(n_{1}n_{2})_{\rm spin1}(s_{3}s_{4})_{\rm spin2}]_{\rm spin3}(c_{5}c_{6})_{\rm spin4}\rangle_{\rm spin5}$ to describe the spin wave functions, 
where spin1, spin2, spin3, spin4, and spin5 represent the spins of $n_{1}n_{2}$, $s_{3}s_{4}$, $n_{1}n_{2}s_{3}s_{4}$ and $c_{5}c_{6}$ and the total spin, respectively.
In Table \ref{base},  $S(A)$ means totally symmetric(antisymmetric) for $nn$, $ss$, and $cc$ of the discussed $nnsscc$ states.

%%%%%%%%%%%%%%%%%%%%%%%%%%%%%%%%%%%%%%%%%%%%%%%%%%%%%%%%%%%%%%%%%%%%
\section{Numerical results and discussion}
\label{sec4}
%%%%%%%%%%%%%%%%%%%%%%%%%%%%%%%%%%%%%%%%%%%%%%%%%%%%%%%%%%%%%%%%%%
By constructing all the possible $F_{flavor} \otimes \phi_{color} \otimes\chi_{spin}$ bases satisfied for $\{1234\}\{56\}$, $\{123\}4\{56\}$, and $\{12\}\{34\}\{56\}$ symmetries and by introducing the parameters collected in Table \ref{para},
we show the obtained corresponding mass spectra for the studied doubly charmed dibaryon states in Table \ref{mass}.
Here, we use the notation $H_{cc,xs}(I, J^{P}, {\rm Mass})$  to label a particular hexaquark state, where $x (x=0, 1, 2, 3, 4)$ represents the number of $s$ quark.

\begin{table*}[t]
\centering \caption{The masses for the dibaryon states $nnnncc$, $nnnscc$, $nnsscc$, $sssncc$, and $sssscc$ in units of MeV. $I$, $J$, and $P$ represent the isospin, angular momentum, and parity, respectively.}
\label{mass}
\begin{lrbox}{\tablebox}
\renewcommand\arraystretch{1.2}
\renewcommand\tabcolsep{2.3pt}
\begin{tabular}{c|cc|c|cc|ccc|c|cccc}
\bottomrule[1.5pt]
\bottomrule[0.5pt]
Dibaryon&$I(J^P)$&Mass&Dibaryon&$I(J^P)$&\multicolumn{1}{c}{Mass}&\multicolumn{1}{c}{$I(J^P)$}&Mass&&Dibaryon&$I(J^P)$&Mass\\
\bottomrule[0.7pt]
\multirow{22}*{$nnnncc$}&$2(2^{+})$&5098&\multirow{14}*{$nnnscc$}&$\frac32(3^{+})$&\multicolumn{1}{c}{5042}&$\frac12(3^{+})$&5093&&
\multirow{29}*{$nnsscc$}&$1(3^{+})$&5209\\
&$2(1^{+})$&5140&&\multirow{3}*{$\frac32(2^{+})$}&\multicolumn{1}{c}{\multirow{3}*{$\begin{pmatrix}5215\\5032\\4940\end{pmatrix}$}}
&\multirow{8}*{$\frac12(2^{+})$}&\multirow{8}*{$\begin{pmatrix}5199\\5159\\5099\\5069\\5025\\4986\\4944\\4838\end{pmatrix}$}
&&&\multirow{5}*{$1(2^{+})$}&\multirow{5}*{$\begin{pmatrix}5327\\5284\\5197\\5105\\5028\end{pmatrix}$}\\
&\multirow{3}*{$2(0^{+})$}&\multirow{3}*{$\begin{pmatrix}5374\\5133\end{pmatrix}$}&&&\multicolumn{1}{c}{}&&&&\\
&&&&&\multicolumn{1}{c}{}&&&&&&&\\
&&&&\multirow{7}*{$\frac32(1^{+})$}&\multicolumn{1}{c}{\multirow{7}*{$\begin{pmatrix}5272\\5260\\5066\\5002\\4945\\4944\\4930\end{pmatrix}$}}&&&&&\\
\Xcline{2-3}{0.5pt}
&$1(3^{+})$&4936&&&\multicolumn{1}{c}{}&&&&&&\\
&\multirow{3}*{$1(2^{+})$}&\multirow{3}*{$\begin{pmatrix}4918\\4795\end{pmatrix}$}&&\multicolumn{1}{c}{}&\multicolumn{1}{c}{}&&&&&\multirow{6}*{$1(1^{+})$}&\multirow{6}*{$\begin{pmatrix}5370\\5335\\5161\\5123\\5064\\4981\end{pmatrix}$}\\
&&&&&\multicolumn{1}{c}{}&\multicolumn{1}{c}{}&&&&\\
&&&&&\multicolumn{1}{c}{}&\multicolumn{1}{c}{}&&&&\\
&\multirow{5}*{$1(1^{+})$}&\multirow{5}*{$\begin{pmatrix}4884\\4807\\4747\\4918\end{pmatrix}$}&&&\multicolumn{1}{c}{}&\multirow{15}*{$\frac12(1^{+})$}&\multirow{15}*{$\begin{pmatrix}5234\\5094\\5063\\5041\\5026\\5022\\5011\\4989\\4986\\4953\\4945\\4908\\4899\\4771\\4603^*\end{pmatrix}$}&&&&\\
&&&&&\multicolumn{1}{c}{}&\multicolumn{1}{c}{}&&&&\\
&&&&\multirow{3}*{$\frac32(0^{+})$}&\multicolumn{1}{c}{\multirow{3}*{$\begin{pmatrix}5456\\5243\\4986\end{pmatrix}$}}&&&&&\\
&&&&&\multicolumn{1}{c}{}&&&&&\multirow{5}*{$1(0^{+})$}&\multirow{5}*{$\begin{pmatrix}5550\\5350\\5203\\5140\\4942\end{pmatrix}$}\\
&&&&&\multicolumn{1}{c}{}&&&&&\\
&\multirow{3}*{$1(0^{+})$}&\multirow{3}*{$\begin{pmatrix}4824\\5107\end{pmatrix}$}&&\multicolumn{1}{c}{}&\multicolumn{1}{c}{}&&&&&\\
\Xcline{4-6}{0.8pt}
&&&\multirow{14}*{$sssncc$}&$\frac12(3^{+})$&5368&&&&&\\
&&&&\multirow{3}*{$\frac12(2^{+})$}&\multirow{3}*{$\begin{pmatrix}5438\\5352\\5237\end{pmatrix}$}&&&&&\\
\Xcline{2-3}{0.5pt}\Xcline{11-12}{0.5pt}
&\multirow{3}*{$0(2^{+})$}&\multirow{3}*{$\begin{pmatrix}5045\\4683\end{pmatrix}$}&&&&&&&&$0(3^{+})$&5242\\
&&&&&&&&&&\multirow{3}*{$0(2^{+})$}&\multirow{3}*{$\begin{pmatrix}5224\\5097\\4904^* \end{pmatrix}$}\\
&&&&\multirow{7}*{$\frac12(1^{+})$}&\multirow{7}*{$\begin{pmatrix}5489\\5460\\5307\\5261\\5196\\5193\\5165\end{pmatrix}$}&&&&\\
&$0(1^{+})$&4635&&&&&&&&&\\
&\multirow{3}*{$0(0^{+)}$}&\multirow{3}*{$\begin{pmatrix}4893\\4577\end{pmatrix}$}&&&&&&&&\multirow{7}*{$0(1^{+})$}
&\multirow{7}*{$\begin{pmatrix}5330\\5181\\5114\\5061\\5025\\4892\\4804^* \end{pmatrix}$}\\
&&&&&&&&&&\\
&&&&&&&&&&\\
\Xcline{1-3}{0.8pt}
\multirow{4}*{$sssscc$}&$0(2^{+})$&5557&&&&\multirow{5}*{$\frac12(0^{+})$}&\multirow{5}*{$\begin{pmatrix}5069\\5027\\4973\\4944\\4755\end{pmatrix}$}&&\\
&$0(1^{+})$&5604&&&&&&&\\
&\multirow{3}*{$0(0^{+})$}&\multirow{3}*{$\begin{pmatrix}5766\\5574\end{pmatrix}$}&&\multirow{3}*{$\frac12(0^{+})$}&\multirow{3}*{$\begin{pmatrix}5652\\5458\\5269\end{pmatrix}$}&&&&&\\
&&&&&&&&&\\
&&&&&&&&&&\multirow{2}*{$0(0^{+})$}
&\multirow{2}*{$\begin{pmatrix}5121\\4827^* \end{pmatrix}$}\\
&&&&&&&&&&\\
\bottomrule[0.5pt]
\bottomrule[1.5pt]
\end{tabular}
\end{lrbox}\scalebox{1.05}{\usebox{\tablebox}}
\end{table*}
%%%%%%%%%%%%%%%%%%%%%%%%%%%%%%%%%%%%%%%%%%%%%%%%%%%%%%%%%%%%%%%%%%%%
\subsection{The $nnnncc$ states}
%%%%%%%%%%%%%%%%%%%%%%%%%%%%%%%%%%%%%%%%%%%%%%%%%%%%%%%%%%%%%%%%%%%%

\begin{figure}[t]
\includegraphics[width=9.1cm]{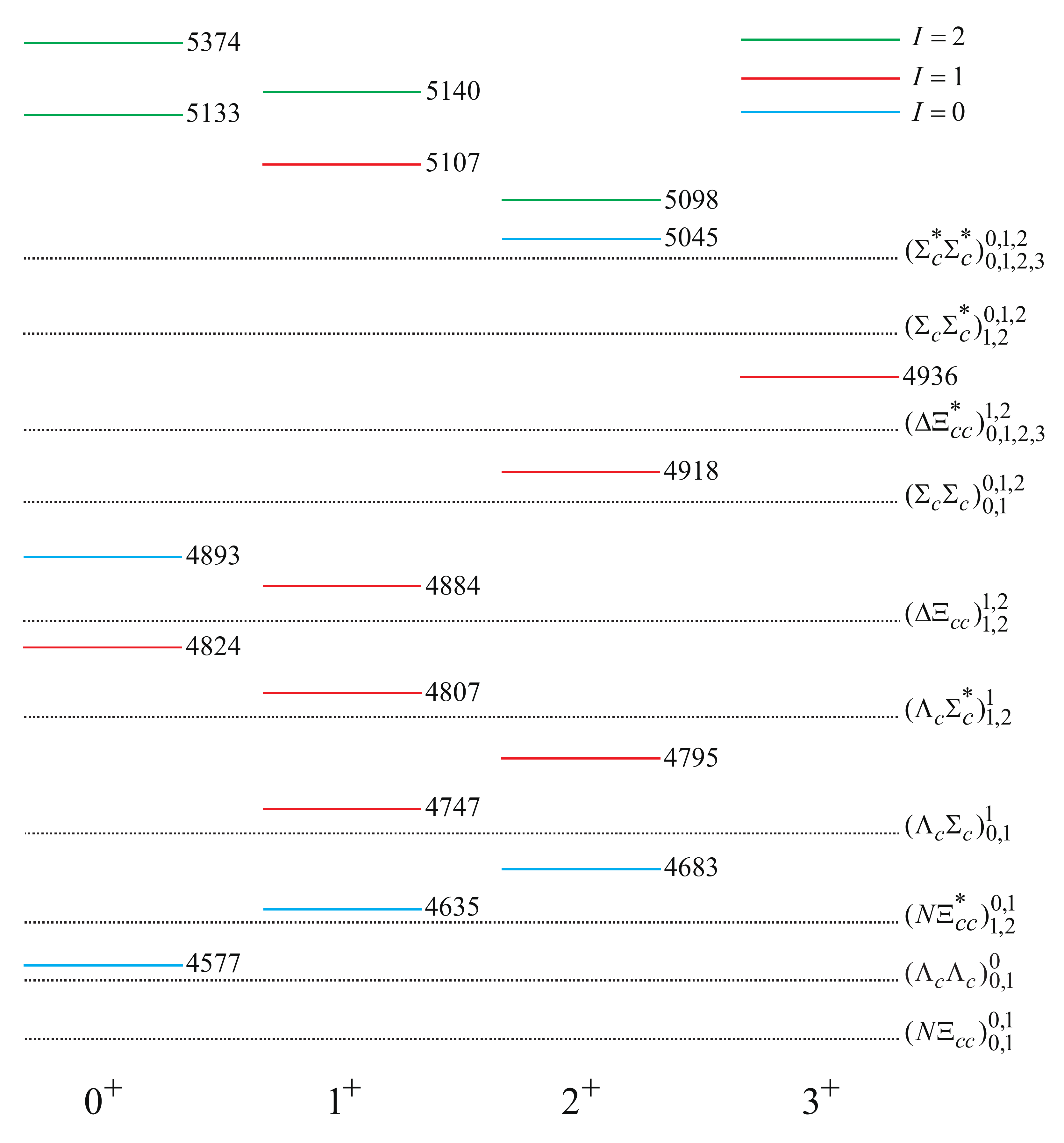}
\caption{Relative positions (units: MeV) for the $nnnncc$ dibaryon states labeled with solid lines.
The green, red, and blue lines represent the $nnnncc$ dibaryon states with $I=2$, $I=1$, and $I=0$, respectively.
The dotted lines denote different baryon-baryon thresholds, and the superscripts (subscripts) of the labels represent the possible total angular momentum (isospin) of the channels. 
	}\label{fig1}
\end{figure}

First, we examine the mass spectrum of the $nnnncc$ ($I=2,1,0$) states. 
Because the four $n$ quarks are identical particles, we choose the wave functions in Sec.~\ref{sec:4ip} as the eigenvector to diagonalize the Hamiltonian matrix.
According to the obtained mass spectrum in Table \ref{mass},
we plot the relative positions for the $nnnncc$ dibaryon states in Fig. \ref{fig1}.
In addition, by rearranging the six constituent quarks, 
we plot the possible thresholds of the baryon-baryon decay patterns for the $nnnncc$ states in Fig. \ref{fig1}.
There are two rearrangement decay types: (single charmed baryon + single charmed baryon) and (light baryon + doubly charmed baryon).
Here, all the possible decay channels in the figure are $\Lambda_{c}\Lambda _{c}$, $\Lambda _{c}\Sigma _{c}$, $\Lambda _{c}\Sigma^{*}_{c}$, $\Sigma _{c}\Sigma _{c}$, $\Sigma^{*}_{c}\Sigma _{c}$, $\Sigma^{*}_{c}\Sigma^{*}_{c}$ and $N\Xi_{cc}$, $N\Xi^{*}_{cc}$, $\Delta\Xi_{cc}$, $\Delta\Xi^{*}_{cc}$. The masses of the ground conventional baryons are taken as the thresholds for these decay channels \cite{Weng:2018mmf}. 
The $nnnncc$ states can decay to the corresponding baryon-baryon final states via the $S$-wave or the $D$-wave. 
According to the conservation law in the strong decay process, the initial states and the final states have the same isospin $I$ and spin-parity $J^P$. 
For convenience, we label the spin (isospin) of the baryon-baryon states with a subscript (superscript) in Fig. \ref{fig1}.
One of these states would be a good $nnnncc$ dibaryon candidate if it were observed in the corresponding decay channels.
 
The positions of these $nnnncc$ states are roughly shown in Fig. \ref{fig1}, and their properties may change accordingly if the masses of $nnnncc$ states are determined by an observed state. Of course, the mass gaps between the different states are reliable.

Because of the complicated mixing between a large number of color-spin wave functions from the strong symmetric constraint, the total number of possible $nnnncc$ states is limited to 17. 
There are 4 isotensor states, 8 isovector states, and 5 isoscalar states.

In Fig. \ref{fig1}, we can see that for the $nnnncc$ states,  
the $J^P=0^{+}$ particles have the smallest and the largest masses among all the quantum numbers.
By labeling the $nnnncc$ states for $I=$2, 1, and 0 with green, red, and blue lines, respectively, 
we can easily see that the $I=0$ $nnnncc$ hexaquarks generally have smaller masses than the $I=1$ states, which have smaller masses than the $I=2$ ones. Thus our results indicate that the states with smaller isospin quantum numbers generally form compact structures easily and have smaller masses.

From Fig. \ref{fig1}, there is no state with $I(J^{P}) = 2(3^+)$ owing to the constraint of the Pauli principle.
Meanwhile, all states with $I=2$ are above and beyond possible baryon-baryon thresholds; thus we think that there is no stable state and these states should have relatively large widths for the isotensor sector. 
In the isovector case, both the heaviest state and the lightest state occur in $J^P=1^+$. 
The mass gap between the two states is about 260 MeV. 
The $H_{cc}(1, 1^{+}, 5107)$ state can decay through all decay channels via the $S$-wave except for $\Lambda_{c} \Lambda_{c}$. 
Similarly, there is no $I(J^P)=0(3^+)$ state, because of the symmetry constraint in the isoscalar sectors. 

According to  Fig. \ref{fig1}, we find that the lowest isoscalar state, i.e.,  
the $H_{cc}(0, 0^{+}, 4577)$ state can only decay into $\Lambda_{c}\Lambda_{c}$ and $N\Xi_{cc}$ final states. 
Furthermore, owing to the conservation of angular momentum and isospin,  
the $H_{cc}(0, 2^{+}, 4683)$ state decays mainly into $N\Xi^{*}_{cc}$ via an $S$-wave, 
but it can also decay into $\Lambda_{c}\Lambda_{c}$ and $N\Xi_{cc}$ via the $D$-wave.
Furthermore, because of the small phase space, this state is expected to be a narrow state.

\subsection{The $nnnscc$ states}
%%%%%%%%%%%%%%%%%%%%%%%%%%%%%%%%%%%%%%%%%%%%%%%%%%%%%%%%%%%%%%%%%%%%

\begin{figure}[t]
	\includegraphics[width=9.1cm]{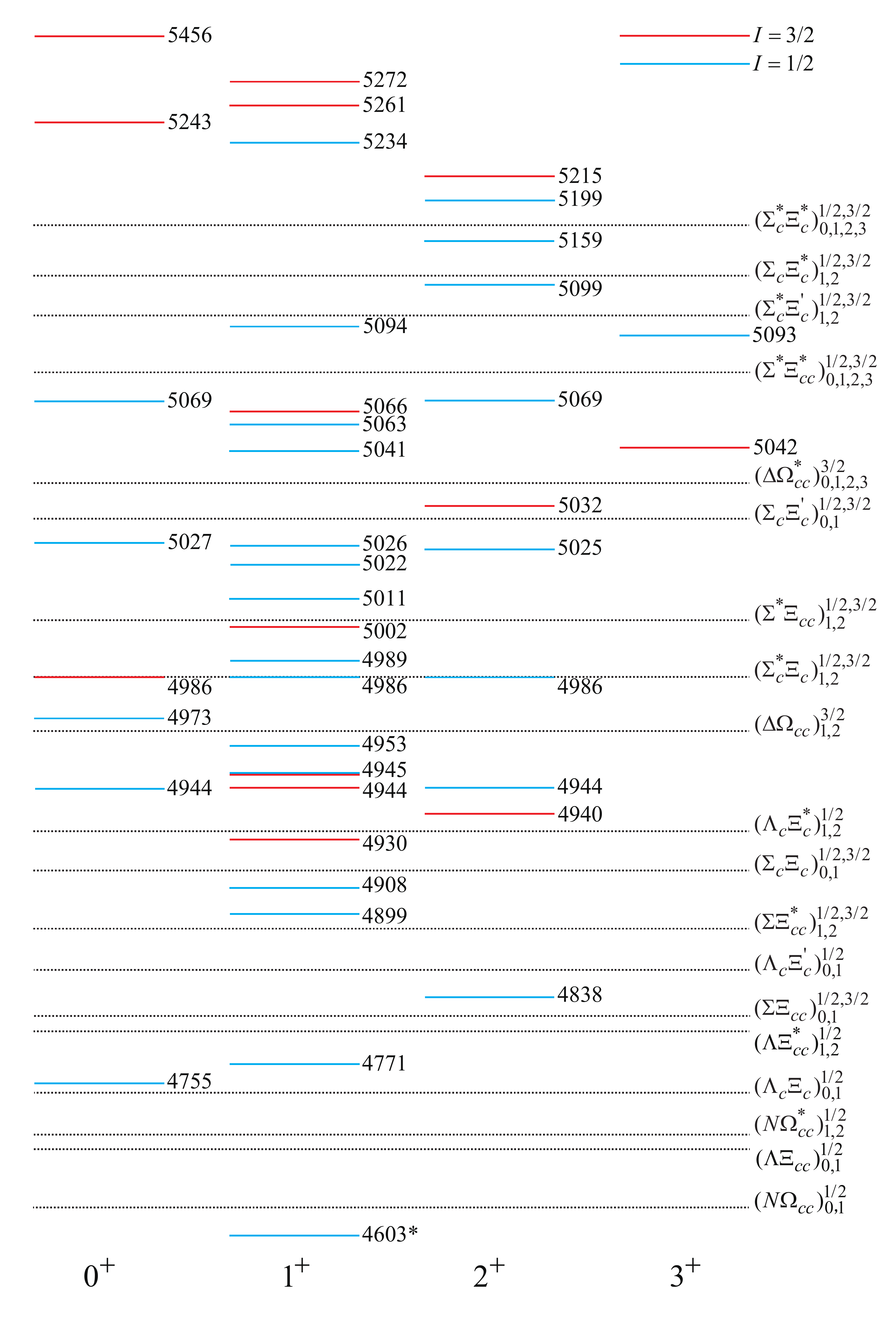}
	\caption{Relative positions (units: MeV) for the $nnnscc$ dibaryon states.
	 The red lines and the blue lines show the $nnnscc$ dibaryon states with $I=3/2$ and $I=1/2$, respectively.
	 The dotted lines denote different baryon-baryon thresholds, and the superscripts (subscripts) of the labels represent the possible total angular momentum (isospin) of the channels. 
	}\label{fig2}
\end{figure}

Next, we investigate the $nnnscc$ ($I=3/2, 1/2$) dibaryon states.
Because three $n$ quarks are identical particles, we choose the wave functions in Sec.~\ref{sec:3ip} as the eigenvector to diagonalize the Hamiltonian matrix.
In contrast to the wave functions of the $nnnncc$ state, the color and spin wave functions are constructed in the ``$baryon~\otimes~baryon$" configuration.
According to the mass spectrum of $nnnscc$ states in Table \ref{mass}, we similarly plot the relative positions for $nnnscc$ states and all the baryon-baryon thresholds of the associated  rearrangement decay patterns in Fig. \ref{fig2}. 
The decays can occur via the $S$- or $D$-wave interactions, and each $nnnscc$ state can decay into these channels from the parity conservation, isospin conversation, and angular momentum conservation.
There are three types of rearrangement decays: ($nnn+ccs$), ($nns+ccn$), and ($cnn+csn$), respectively. 

In Fig. \ref{fig2}, we label the $nnnscc$ states for $I=3/2$ and $1/2$ with red and blue lines, respectively.
We can easily find that the $I=1/2$ states generally have smaller masses than the $I=3/2$ $nnnscc$ states. 
Thus, our results indicate that the states with smaller isospin quantum numbers generally form compact structures easily and have smaller masses.
From Fig. \ref{fig2}, we find that the lowest $I(J^{P})=1/2(1^{+})$ $H_{cc,s}(4603, 1/2, 1^{+})$ state
is below all the rearrangement decay channels, and its decay via baryon-baryon channels is kinetically forbidden.
We therefore consider it to be a relatively stable state. 
We label this stable state with ``$\ast$'' in Fig. \ref{fig2} and Table \ref{mass}.

In the $nnnscc$ $(I=3/2)$ case, most states are expected to have relatively large widths, because they all have many different rearrangement decay channels.
For example, consider the lowest $I(J^{P})=3/2(2^{+})$ state: $H_{cc,s}(3/2, 2^{+}, 4940)$ can easily decay to  $\Sigma\Xi^{*}_{cc}$ final states via the $S$ wave.
It also decays into $\Sigma_{c}\Xi_{c}$, $\Sigma\Xi_{cc}$ final states by via the $D$ wave.

As for the $nnnscc$ states with $I=1/2$, most of them are unstable particles against strong decay. 
Both the heaviest state and the lightest states occur in $J^{P}=1^+$. 
Here, the $H_{cc,s}(1/2, 0^{+}, 4755)$ and $H_{cc,s}(1/2, 1^{+}, 4771)$ states have the same three quark-rearrangement decay channels
$\Lambda_{c}\Xi_{c}$, $\Lambda\Xi_{cc}$, and $N\Omega_{cc}$.
Moreover, the $H_{cc,s}(1/2, 3^{+}, 5093)$ state has only one baryon-baryon decay channel $\Sigma^{*}\Xi^{*}_{cc}$ via the $S$ wave.
Meanwhile, it has many different decay channels via the $D$ wave.

%%%%%%%%%%%%%%%%%%%%%%%%%%%%%%%%%%%%%%%%%%%%%%%%%%%%%%%%%%%%%%%%%%%%
\subsection{The $nnsscc$ states}
%%%%%%%%%%%%%%%%%%%%%%%%%%%%%%%%%%%%%%%%%%%%%%%%%%%%%%%%%%%%%%%%%%%%

 For dealing with the permutation symmetry of identical quarks $n$-$n$, $s$-$s$, and $c$-$c$, the wave functions are presented in Sec.~\ref{sec:2ip}.
The flavor, color, and spin wave functions are constructed in the $\left | (nn)\otimes (ss)\otimes (cc) \right \rangle$ configuration.
According to the corresponding mass spectrum in Table \ref{mass},
we plot the relative positions for the $nnsscc$ dibaryon states in Fig. \ref{fig3} (a).
Meanwhile, all the possible baryon-baryon thresholds of the associated rearrangement decay patterns are shown in Fig. \ref{fig3} (a).
There are four types of rearrangement decays: ($nns+scc$), ($nss+ncc$), ($nsc+nsc$), and ($nnc+ssc$).

By labeling the $nnsscc$ states for $I=1$ and $0$ with red and blue lines, respectively, 
It is easy to see that the $I=0$ states generally have smaller masses than the $I=1$ states.
Furthermore, it is easy to see that the two lowest $nnccss$ dibaryon states, i.e., $H_{cc,2s}(4827,0,0^{+})$ and $H_{cc,2s}(4804, 0, 1^{+})$, should be stable, which are below all the baryon-baryon thresholds, and these two states are probably narrow.

For the isovector $nnsscc$ states, the heaviest state and the lightest state are both the $J^P=0^+$ states, i.e., $H_{cc,2s}(1, 0^+, 5550)$ and $H_{cc,2s}(1, 0^+,4942)$. The mass gap between the two states is about 610 MeV. All the states are expected to be broad states with many different rearrangement decay channels.

For the isoscalar $nnsscc$ states, except for the relatively stable states, most of them have many different decay channels.
Meanwhile, the $H_{cc,2s}(4892,0,1^{+})$ state is lower than the other thresholds and just above the $\Lambda\Omega_{cc}$ threshold; thus it is probably narrow, and we can use the $\Lambda\Omega_{cc}$ channel to identify this state. 
Furthermore, the lowest $I(J^{P})=0(2^{+})$ state, i.e., $H_{cc,2s}(4904,0,2^{+})$, can only decay into $\Lambda\Omega_{cc}$ via a $D$ wave; thus, it should be narrow.

%%%%%%%%%%%%%%%%%%%%%%%%%%%%%%%%%%%%%%%%%%%%%%%%%%%%%%%%%%%%%%%%%%%%
\subsection{The $sssncc$ and $sssscc$ states}
%%%%%%%%%%%%%%%%%%%%%%%%%%%%%%%%%%%%%%%%%%%%%%%%%%%%%%%%%%%%%%%%%%%%

\begin{figure*}[t]
	\includegraphics[width=18.1cm]{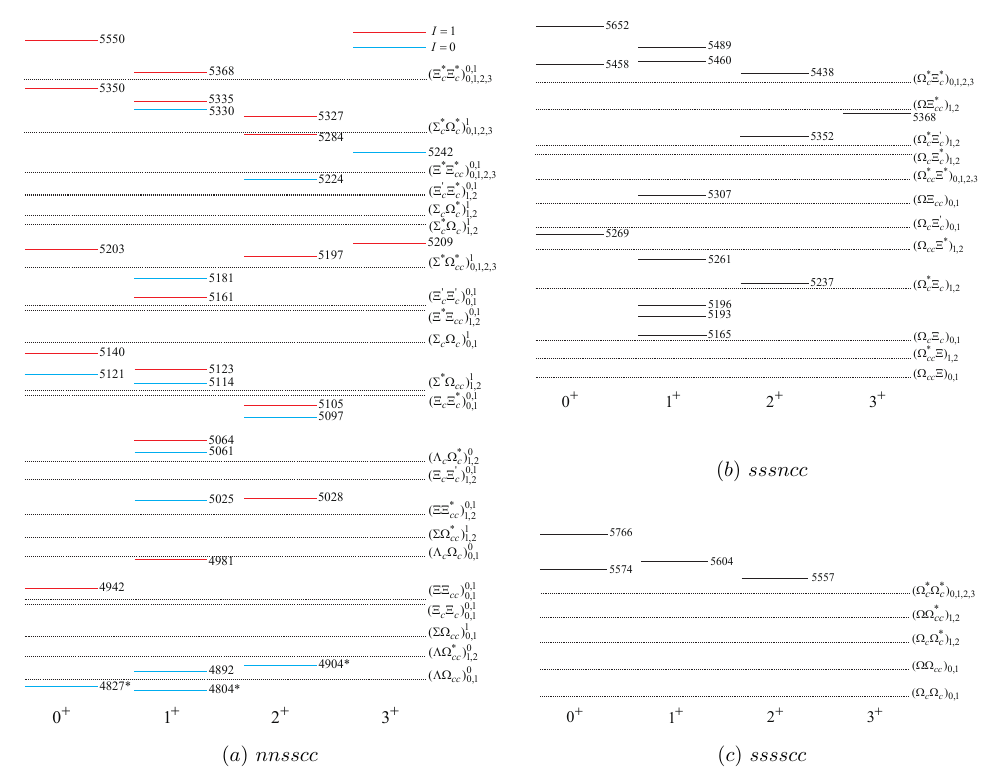}
	\caption{Relative positions (units: MeV) for three types of hexaquark states.
		(a) The $nnsscc$ subsystem:
		the red lines and the blue lines show the $nnsscc$ hexaquark states with $I=1$ and $I=0$ respectively;\label{a}
		(b) the $sssncc$ subsystem;\label{b} (c) the $sssscc$ subsystem.\label{c}
		The dotted lines denote different baryon-baryon thresholds, and the superscripts (subscripts) of the labels represent the possible total angular momentum (isospin) of the channels. https://www.overleaf.com/project/638ca57a6d52186122fd3861
	}\label{fig3}
\end{figure*}

Finally,  we discuss the $sssncc$ and $sssscc$ dibaryon states. 
The number of these allowed $sssncc$ dibaryon states is the same as the number of  $nnnscc$ $(I=3/2)$ dibaryon states, owing to the same symmetry requirements according to Table \ref{mass}. 
Similarly, the number of allowed $sssscc$ dibaryon states is the same as the number of $nnnncc$ $(I=2)$ dibaryon states.
Furthermore, the relative positions for the $sssncc$ and $sssscc$ dibaryon states and their all possible rearrangement decay channels are shown in Fig. \ref{c} (b) and (c), respectively.

The $sssncc$ dibaryon states have 12 possible rearrangement decay channels:  $\Omega^{*}_{c}\Xi^{*}_{c}$, $\Omega^{*}_{c}\Xi'_{c}$, 
$\Omega^{*}_{c}\Xi_{c}$,  $\Omega_{c}\Xi^{*}_{c}$, $\Omega_{c}\Xi'_{c}$, $\Omega_{c}\Xi_{c}$,  $\Omega\Xi^{*}_{cc}$, $\Omega\Xi_{cc}$, $\Omega^{*}_{cc}\Xi^{*}$,  $\Omega^{*}_{cc}\Xi$,
$\Omega_{cc}\Xi^{*}$, and $\Omega_{cc}\Xi$.
The $sssscc$ dibaryon states, they have 5 possible rearrangement decay channels: $\Omega^{*}_{c}\Omega^{*}_{c}$, $\Omega\Omega^{*}_{cc}$, $\Omega_{c}\Omega^{*}_{c}$, $\Omega\Omega_{cc}$, and
$\Omega_{c}\Omega_{c}$. From Fig. \ref{c} (c), we notice that the ground $sssscc$ dibaryon state with quantum number $I(J^{P})=0(3^{+})$
does not exist, owing to the constraint of the Pauli principle.

From Fig. \ref{fig3} (b), all the states are above some possible baryon-baryon thresholds in the $sssncc$ dibaryon state.
Thus, they should have relatively large widths.
As expected, all the $sssscc$ dibaryon states are above the allowed baryon-baryon thresholds from Fig. \ref{fig3} (c).
They can easily decay into the rearrangement channels owing to the large phase space. 
Accordingly, all the $sssscc$ dibaryon states are likely to be broad. It does not seem easy to find them experimentally.

%%%%%%%%%%%%%%%%%%%%%%%%%%%%%%%%%%%%%%%%%%%%%%%%%%%%%%%%%%%%%%%%%%%%
\section{SUMMARY}
%%%%%%%%%%%%%%%%%%%%%%%%%%%%%%%%%%%%%%%%%%%%%%%%%%%%%%%%%%%%%%%%%%%%
\label{sec5}
In 2021, the LHCb Collaboration reported the first doubly charmed tetraquark state $T^{+}_{cc}$ ($cc\bar{u}\bar{d}$). Inspired by this discovery, the existence of doubly heavy tetraquark states or pentaquark states with different quark configurations has been studied via various theoretical calculations. According to the similarity between the $cc\bar{u}\bar{d}$ tetraquark state and the $\Xi^{++}_{cc}$ baryon ($ccu$), we naturally ask whether the doubly charmed dibaryon states exist. Simultaneously, it is an interesting question whether the dibaryon state with double heavy quarks is also stable against strong decay. 

In this work, we first introduce the CMI model and construct the $F_{flavor} \otimes \phi_{color} \otimes\chi_{spin}$ wave functions for the doubly charmed dibaryon system.
The main challenge here is how to apply the Pauli principle to the permutation of the three or four identical particles. 
Therefore, we introduce the Young tableau to represent the irreducible bases of the permutation group. 
Then, we systematically calculate the corresponding Hamiltonian. 
Furthermore, we obtain the mass spectra of the dibaryon states $nnnncc$, $nnnscc$, $nnsscc$, $sssncc$, and $sssscc$, 
which are listed in Table \ref{mass}. 
Accordingly, the relative positions between dibaryon states and possible baryon-baryon thresholds are plotted in Figs. \ref{fig1}-\ref{fig3}.
Through comparison with baryon-baryon thresholds, we can analyze the stability of dibaryon states and the decay properties of the possible quark rearrangement decay channels.
Furthermore, the mass gaps between the relevant states are reliable, and if a $qqqqcc$ dibaryon state is observed, we can use these mass gaps to predict its corresponding multiples.

According to the obtained mass spectrum, there is no state with $I(J^{P})=2(3^{+})$ and $0(3^{+})$ among the $nnnncc$ and $sssscc$ states, owing to the constraint of the Pauli principle.
Moreover, our results show that the states with smaller isospin quantum numbers generally form a compact structure easily and have smaller masses in the $nnnncc$ ($I=2, 1, 0$), $nnnscc$ ($I=3/2, 1/2$), and $nnsscc$ ($I=1, 0$) states.
Meanwhile, we find four relatively ``stable" states: $H_{cc,s}(1/2, 1^{+}, 4603)$, $H_{cc,2s}(0, 0^+, 4827)$, and $H_{cc,2s}(0, 1^+, 4804)$, whose decays through baryon-baryon channels are kinetically forbidden.

\begin{table*}[t]
\centering \caption{Some possible weak decay modes of the three stable hexaquark states.
}\label{decay}
\begin{lrbox}{\tablebox}
\renewcommand\arraystretch{1.5}
\renewcommand\tabcolsep{5pt}
\begin{tabular}{c|c}
\bottomrule[1.5pt]
\bottomrule[0.5pt]
Hexaquark states & Weak decay modes\\
\bottomrule[0.5pt]
$H_{cc,s}(1/2, 1^{+}, 4603)$ &  $\Sigma^{+}_{c}pK^{-}(\pi^{-})$, $\Xi^{+}_{c}\Sigma^{-}\rho^{+}(\pi^{+})$, $\Lambda_{c}^{+}\Sigma^{+}K^{-}$, $\Sigma_{c}^{+}\Sigma^{+}K^{-}$, $\Xi_{c}^{+}pK^{+}K^{-}$, $\Xi_{c}^{+}\Delta^{++}K^{-}$, ......          \\ 
\bottomrule[0.5pt]
$H_{cc,2s}(0, 0^+, 4827)$    & \multirow{2}{*}{$\Xi^{+}_{c}\Sigma^{+}K^{-}$, $\Xi^{+}_{c}\Xi^{-}\pi^{+}$, $\Xi^{+}_{c}\Xi^{-}\rho^{+}(\pi^{+})$, $\Lambda_{c}^{+}\Omega^{-}\rho^{+}(\pi^{+})$, ......}            \\ 
$H_{cc,2s}(0, 1^+, 4804)$    &                     \\
\bottomrule[0.5pt]
\bottomrule[1.5pt]
\end{tabular}
\end{lrbox}\scalebox{1.08}{\usebox{\tablebox}}
\end{table*}

Hence, we can explore the presence of the three ``stable" double charm hexaquark states by investigating their weak decay channels. We assume that only one charm quark decays into $d\bar d u$, $s\bar d u$ or $s\bar s u$ \cite{Workman:2022ynf, LHCb:2017iph, Dhir:2018twm, Solovieva:2008fw, E687:1992tqn}, and all quarks can form various hadron final states through quark rearrangement. However, the weak decay process must obey the laws of conservation of charge and angular momentum. In Table~\ref{decay}, we list some possible weak decay modes for the three ``stable" states, focusing on electric hadron final states, which could aid in experimental reconstruction of the hexaquark state. For instance, in $nnnscc$, the $H_{cc,s}(1/2, 1^{+}, 4603)$ states carry one or two units of positive charges and may decay into $\Sigma^{+}_{c}pK^{-}(\pi^{-})$, $\Xi^{+}_{c}\Sigma^{-}\rho^{+}(\pi^{+})$, $\Lambda_{c}^{+}\Sigma^{+}K^{-}$, $\Sigma_{c}^{+}\Sigma^{+}K^{-}$, $\Xi_{c}^{+}pK^{+}K^{-}$, $\Xi_{c}^{+}\Delta^{++}K^{-}$, etc. Similarly, the two states $H_{cc,2s}(0, 0^+, 4827)$ and $H_{cc,2s}(0, 1^+, 4804)$ share common decay modes such as $\Xi^{+}_{c}\Sigma^{+}K^{-}$, $\Xi^{+}_{c}\Xi^{-}\pi^{+}$, $\Xi^{+}_{c}\Xi^{-}\rho^{+}(\pi^{+})$ and $\Lambda_{c}^{+}\Omega^{-}\rho^{+}(\pi^{+})$. The weak decay modes presented in Table~\ref{decay} correspond to final states containing two electric baryons and one electric meson. Theoretically more electric or neutral mesons could appear in the decay final states.

Of necessity, owing to the uncertainty of the CMI model, further dynamical calculations are needed to clarify their nature.
We hope that the present study will inspire relevant experiments to search for these states in the future.

\section*{Acknowledgements}

This work is supported by the China National Funds for Distinguished Young Scientists under Grant No. 11825503, the National Key Research and Development Program of China under Contract No. 2020YFA0406400, the 111 Project under Grant No. B20063, the National Natural Science Foundation of China under Grant Nos. 12175091, 11965016, 12247101 and 12047501, and the project for top-notch innovative talents of Gansu province.

\end{document}